\documentclass[
 reprint,
 aps,prl,
 amsmath,amssymb,
 showpacs,
]{revtex4-1}

\usepackage{graphicx}  
\usepackage{dcolumn}   
\usepackage{bm}		   
\usepackage{units}	   

\newcommand*{\chem}[1]{$\mathrm{#1}$}

\begin{document}


\title{DNA twist stability changes with magnesium(2+) concentration}

\author{Onno D. Broekmans}
\author{Graeme A. King}
\author{Greg J. Stephens}
\author{Gijs J.L. Wuite}
  \email[Corresponding author. ]{g.j.l.wuite@vu.nl}
 
\affiliation{LaserLaB Amsterdam and Department of Physics and Astronomy\\
VU University Amsterdam, 1081 HV Amsterdam, The Netherlands}

\date{\today}

\begin{abstract}
For an understanding of DNA elasticity at high mechanical loads ($F > \unit[30]{pN}$), its helical nature needs to be taken into account, in the form of coupling between the twist and stretch degrees of freedom. The prevailing model for DNA elasticity, the worm-like chain, was previously extended to include this twist--stretch coupling, giving rise to the twistable worm-like chain. Motivated by DNA's charged nature, and the known effects of ionic charges on the molecule's persistence length and stretch modulus, we explored the impact of buffer ionic conditions on twist--stretch coupling. After developing a robust fitting approach for force--extension data, we find that DNA's helical twist is stabilized at high concentrations of the magnesium divalent cation.
\end{abstract}

    
\maketitle



\begin{figure}
\includegraphics{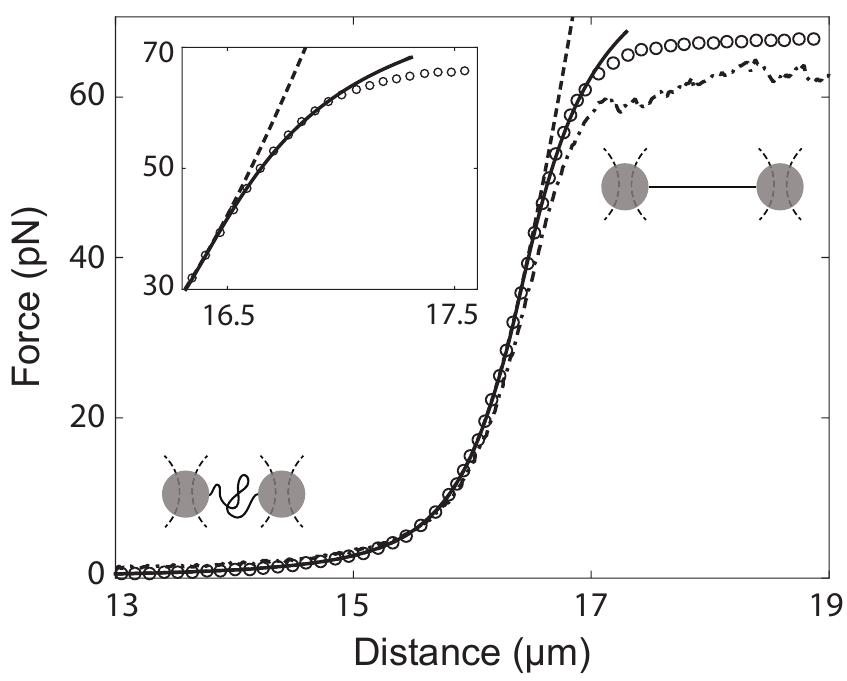}
\caption{\label{fig:typical data + fits} Typical force--extension data for double-stranded DNA ($\circ$) stretched using optical tweezers (schematically shown: a DNA molecule is tethered between two microbeads held in two optical traps). Whereas the extensible worm-like chain ($-\,-$) only fits the data for forces up to $\sim\unit[30]{pN}$, the twistable worm-like chain (---) expands this range to $\sim\unit[60]{pN}$. Shown is averaged data for buffer conditions ($\unit[500]{mM}$ \chem{NaCl}, no \chem{Mg^{2+}}, $N=287$). For comparison, a low-salt curve ($\unit[75]{mM}$ \chem{NaCl}, $-\cdot$) is shown, illustrating the effects of end-unpeeling.}
\end{figure}

Mechanical perturbations of the DNA double helix form a crucial step during many of the cell's life-sustaining processes. When proteins bind, replicate, compact, and repair the genome, the DNA molecule is bent, stretched, and twisted. A detailed understanding of DNA's elastic response to these perturbations is therefore a prerequisite for a deep quantitative insight into the biology of the cell.
The single-molecule techniques that have been developed over the past two decades \cite{Smith1992,Wang1997} have greatly contributed to this understanding: it is now routinely possible to directly manipulate single molecules of DNA, and monitor their response to stretch and twist under a wide variety of experimental conditions.
One such technique is the optical tweezers (FIG.~\ref{fig:typical data + fits}, schematic), which can be used to accurately measure the force response of DNA \cite{Moffitt2008}.
By modeling the corresponding force--extension data, we can then not only improve upon our structural understanding of DNA; it is also possible to infer the mechanisms of action of DNA-binding proteins from the changes they induce in force--extension curves \cite{Heller2014}.

\begin{figure}
\includegraphics[width=6cm]{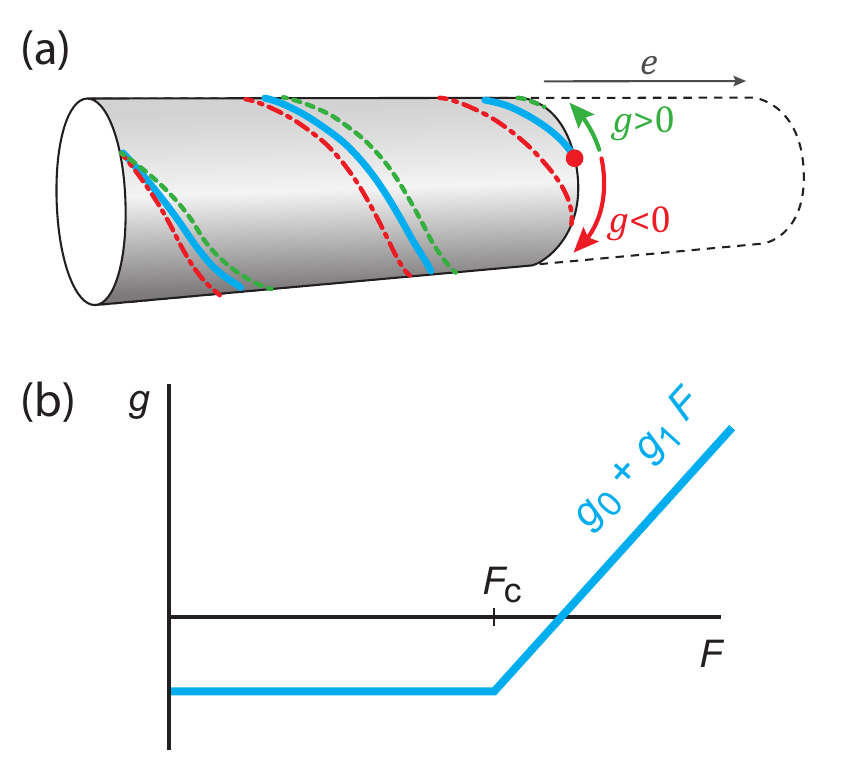}
\caption{\label{fig:twlc model} (Color online) (a) A ``toy model'' of DNA twist--stretch coupling, based on \cite{Gore2006}. For positive values of the twist--stretch coupling $g$, the right-handed DNA helix (blue, ---) would unwind when stretched (green, $-\,-$). In reality, for forces up to $\sim\unit[35]{pN}$, DNA \emph{overwinds} when stretched (red, $-\cdot$). (b) The force dependence of the twist--stretch coupling $g$ of dsDNA is modeled in the twistable worm-like chain as a piecewise linear function: constant up to a critical force $F_c$ ($\unit[30.6]{pN}$ \cite{Gross2011}); above $F_c$, a linear approximation $g(F) = g_0 + g_1 F$ is used.}
\end{figure}

Below mechanical loads of $\sim\unit[30]{pN}$, dsDNA's force response is accurately modeled by the extensible worm-like chain (eWLC) \cite{Odijk1995,Marko1995}. This well-established, semiclassical model describes the molecule as simply an isotropic, extensible rod: Entropic bending fluctuations, characterized by a persistence length $L_p$ ($\unit[50]{nm}$ for dsDNA under physiological conditions), and enthalpic stretching of the DNA backbone, characterized by the stretch modulus $S$ ($\unit[1500]{pN}$ \cite{Gross2011}), are balanced by the work performed by the stretching force $F$, leading to a relative extension $e = d/L_c$ (end-to-end distance over contour length). Beyond $\unit[30]{pN}$, however, dsDNA's helical structure needs to be taken into account \cite{Gore2006,Lionnet2006,Gross2011}. This introduces an energy cross-term between the molecule's twist and stretch degrees of freedom. Only relatively recently, Gross \emph{et al.} found it possible to incorporate this effect into the eWLC, yielding the ``twistable worm-like chain'' (tWLC) \cite{Gross2011}:
\begin{equation}\label{eq:tWLC}
\frac{d}{L_c} = 1 - \frac{1}{2}\left(\frac{k_B T}{F L_p}\right)^{1/2} + \frac{C}{-g(F)^2+SC} \cdot F,
\end{equation}
where $C$ is the molecule's twist rigidity ($\unit[440]{pN\,nm^2}$ \cite{Gross2011}), and $g(F)$ is the force-dependent twist--stretch coupling. As illustrated schematically in FIG.~\ref{fig:twlc model}(a), a positive value of $g$ would correspond to DNA unwinding as it is being stretched. In reality, DNA slightly \emph{overwinds} up to $\sim\unit[35]{pN}$ --- a finding with important implications for proteins that have to twist DNA upon binding \cite{Gore2006,Lionnet2006}. Only at higher forces does the molecule begin to unwind \cite{Gore2006,Gross2011}, until, around $\unit[65]{pN}$, it undergoes a structural transition known as overstretching \cite{Bustamante1994,Mameren2009,Gross2011,Zhang2013}.
The change from overwinding to unwinding has been modeled in the tWLC by taking $g$ as a piecewise linear function of the force $F$ [FIG.~\ref{fig:twlc model}(b)].
Overall, the tWLC substantially extends the range of forces over which dsDNA force--extension data is understood (see FIG.~\ref{fig:typical data + fits}).

More importantly, the tWLC captures DNA twist--stretch coupling into the two parameters $g_0$ and $g_1$, which can now be obtained from fits to force--extension data. This opens up the possibility of investigating if, and to what extent, twist--stretch coupling is affected by buffer ionic conditions. DNA is, after all, a highly charged molecule, known to interact strongly with cations through its phosphate backbone and major groove \cite{Chiu2000,Guéroult2012}. These interactions can lead to softening of DNA \cite{Baumann1997,Wenner2002}, and, for multivalent cations, to bending \cite{Jerkovic2001}, and even condensation \cite{Broek2010}. In this Letter, we therefore set out to quantify DNA twist--stretch coupling as a function of the concentration of the divalent magnesium cation (\chem{Mg^{2+}}).

\begin{figure}
\includegraphics[width=6.5cm]{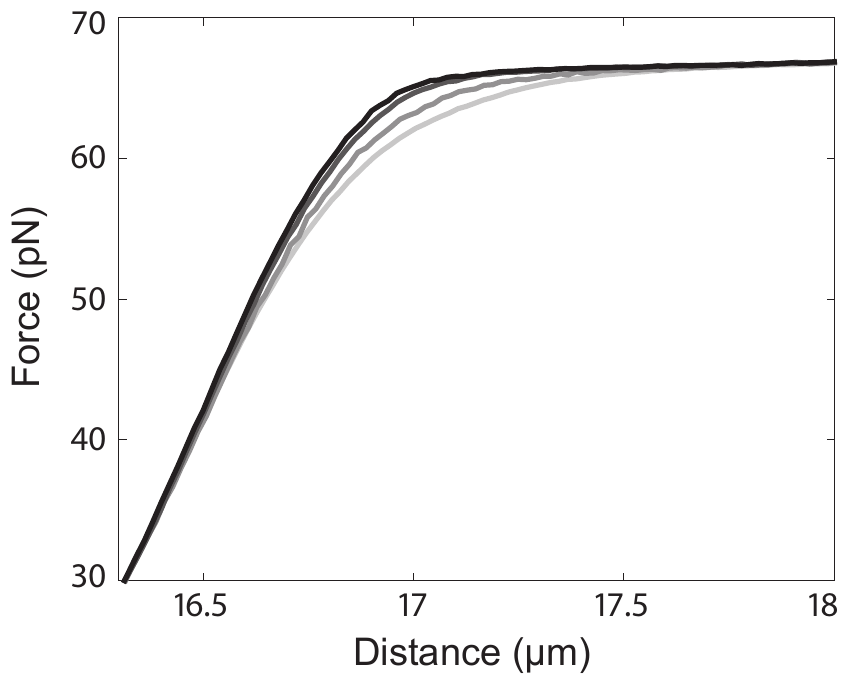}
\caption{\label{fig:stiffening} Double-stranded DNA force--extension data for increasing concentrations of \chem{MgCl_2} (in a background of $\unit[500]{mM}$ \chem{NaCl} and $\unit[10]{mM}$ TrisHCl, pH 7.6; light to dark: 0, 25, 50, and $\unit[100]{mM}$ of \chem{MgCl_2}, respectively). For high magnesium concentrations, a clear stiffening of the DNA before the onset of overstretching is observed. Shown is data averaged per magnesium concentration ($N \geq 12$), after correction for systematic measurement errors (see main text).}
\end{figure}

\paragraph{Optical tweezers experiments.} ---
Using optical tweezers, we collected single-molecule force--extension data on dsDNA at varying concentrations of magnesium (0--$\unit[150]{mM}$ \chem{MgCl_2}). In brief, we tethered biotinylated $\lambda$-phage DNA ($L_c = \unit[16.5]{\mu m}$) between two $\unit[3.05]{{\mu}m}$ streptavidin-coated polystyrene microspheres captured in two optical traps, inside a microfluidic flowcell (for details on the instrument and protocols, please refer to \cite{Candelli2011}). To suppress unpeeling of the untethered ends of the DNA strands, the effects of which are shown in FIG.~\ref{fig:typical data + fits}, we worked in a background of $\unit[500]{mM}$ monovalent salt (\chem{NaCl}) \cite{Gross2011}.
Below forces of $\unit[30]{pN}$, force--extension curves were indistinguishable (data not shown), indicating that the persistence length $L_p$ and stretch modulus $S$ were not affected by the addition of divalent salt.
Since the force dependence of twist--stretch coupling only affects dsDNA elasticity significantly above $\sim\unit[45]{pN}$, we instead focused on comparing the data in this high-force regime (FIG.~\ref{fig:stiffening}). For high concentrations of magnesium, we observed a distinct stiffening of the DNA just before the onset of overstretching. This suggested that twist--stretch coupling was specifically affected.

\paragraph{Data analysis approach.} ---
To quantify the stiffening effect, we fitted the data with the tWLC model, Eq.~(\ref{eq:tWLC}). Since previous reports have shown the twist rigidity $C$ to be insensitive to ionic strength \cite{Mohammad-Rafiee2004,Mosconi2009}, that leaves the model with four free fit parameters: $L_p$, $S$, $g_0$, and $g_1$.
Given this large number of parameters, a solid approach for fitting the data was needed. We would like to highlight three key points in the approach we have developed: (1) fitting with the force as the dependent variable; (2) correcting for systematic measurement errors; and (3) global fitting with shared physical parameters. (For additional details, we refer to the SI, which includes both the raw data, as well as the MATLAB code used for generating the figures in this Letter.)

\begin{figure}
\includegraphics[width=6.5cm]{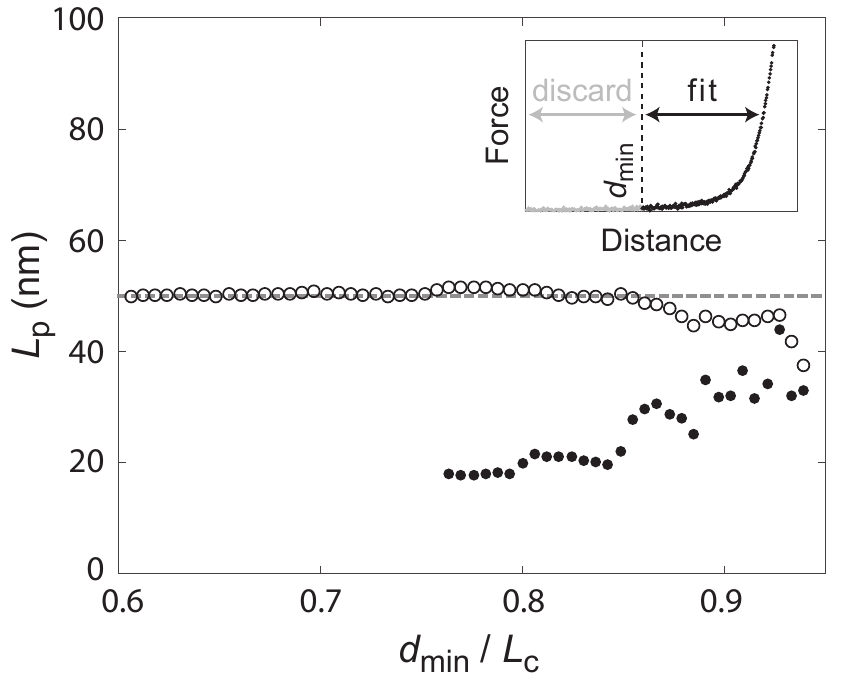}
\caption{\label{fig:wlc eq inversion} When fitting optical tweezers force--extension data with the extensible worm-like chain (eWLC) equation as expressed in Eq.~(\ref{eq:tWLC}), with $d$ as the dependent variable, the fit results are highly dependent on the data range used for fitting. As an illustration of this, fragments of simulated data were fitted using nonlinear least squares, using only the data between a distance $d_{\text{min}}$ and the eWLC's upper limit of $\unit[30]{pN}$ (inset). The values found for the persistence length $L_p$ ($\bullet$) as $d_{\text{min}}$ is varied, significantly underestimate the actual value ($-\,-$), and depend strongly on $d_{\text{min}}$ (missing points indicate a lack of convergence of the fit). When using an \emph{inverted} version of Eq.~(\ref{eq:tWLC}), which expresses $F$ as a function of $d$, a robust and reliable value for $L_p$ is found instead ($\circ$).}
\end{figure}

The first point stems from the observation that, in optical tweezers data, the force signal carries the most significant error (and not the distance, which is precisely controlled). As such, when performing a least-squares fit, the force should be the dependent variable \cite{Bevington2003}. This implies that the model fitted to our data should be an inversion of Eq.~(\ref{eq:tWLC}), expressing force as a function of distance. The impact of this inversion is illustrated in FIG.~\ref{fig:wlc eq inversion}, for the simplified case of an eWLC fit to simulated data. If Eq.~(\ref{eq:tWLC}) is (incorrectly) used as-is for the least-squares fit, the value found for $L_p$ changes wildly as more or less data from the low-force tail is included in the fit --- in addition to systematically underestimating $L_p$. When, instead, an inverted version of Eq.~(\ref{eq:tWLC}) is used, the fit result does become reliable \footnote{We should note that this is a non-issue for magnetic tweezers experiments: in such data, the distance variable carries the most significant error, and Eq.~(\ref{eq:tWLC}) \emph{can} be used as-is.}.

Second of all, we applied corrections for three systematic measurement errors that are intrinsic to our data:
(a) the force at zero extension is not set exactly to zero during each experiment, leading to a random force offset $F_0$ for each force--extension curve;
(b) small variations in microbead diameter lead to a distance offset $d_0$;
and
(c) imperfect force sensor calibration causes force data to include a random factor $\delta F$.
The first two systematic errors were accounted for by including the offsets in the eWLC equation (see SI). This amended eWLC equation was fit to the data below $\unit[30]{pN}$, and the offsets found were subtracted from the data. The third systematic error was rectified by using the force $F_\mathrm{os}$ at which the overstretching plateau occurs as a proxy for $\delta F$. Within the force resolution of our instrument, there appears to be no correlation between magnesium concentration and $F_\mathrm{os}$; we therefore rescaled all force--extension curves to have overlapping overstretching plateaus.

Thirdly, and finally, we opted for a global fitting approach. We grouped all force--extension curves into ensembles by magnesium concentration, implying that the values of the physical parameters (i.e., $L_p$, $S$, $g_0$, and $g_1$) for curves within each ensemble should be equal. We could thus fit all curves in each ensemble simultaneously, while sharing fit parameters between curves. Fits of simulated data confirmed that, generally, global fitting performs significantly better than individual fitting of the curves, with a decreased sensitivity to the aforementioned systematic measurement errors (see SI).

As can be seen in FIG.~\ref{fig:typical data + fits}, the tWLC does not fit the full dsDNA force--extension curve up until the overstretching plateau. We therefore removed all force--extension data above a maximum force $F_\mathrm{max}$, determined by optimizing $F_\mathrm{max}$ in each magnesium concentration ensemble for a maximum goodness-of-fit ($\chi^2$) to the data. This way, we were finally able to determine the tWLC fit parameters for each of the measured magnesium concentrations.

\begin{figure}
\includegraphics{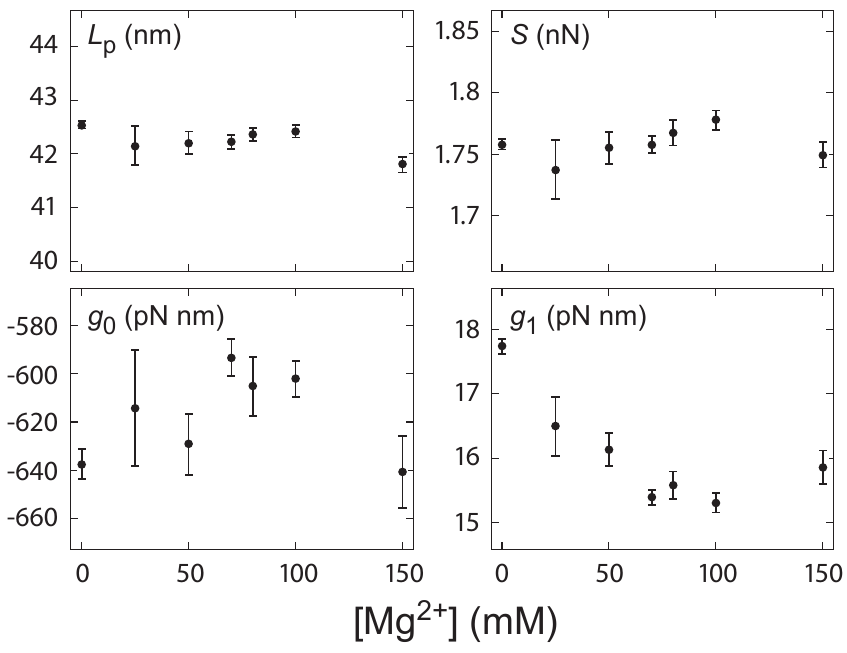}
\caption{\label{fig:twlc results} As the concentration of magnesium(2+) (\chem{MgCl_2}) is increased from 0 to $\unit[150]{mM}$ (in a background of $\unit[500]{mM}$ \chem{NaCl}), the twist--stretch coupling parameter $g_1$ decreases, indicating a stabilization of dsDNA twist (error bars: bootstrap error, $N \geq 12$). In contrast, the persistence length $L_p$ and stretch modulus $S$ are relatively insensitive to these changes in divalent cation concentration.}
\end{figure}

\paragraph{Analysis results.} ---
As shown in FIG.~\ref{fig:twlc results}, the persistence length $L_p$ and the stretch modulus $S$ are stable over the range of magnesium concentrations investigated. This is consistent with the observation that force--extension data below $\unit[30]{pN}$ (the limit of the eWLC) is equal across magnesium concentrations. Indeed, any electrostatic effects on these parameters are expected to be fully saturated well below the monovalent salt concentration ($\unit[500]{mM}$ NaCl) used in our buffer \cite{Baumann1997}.

$g_1$, on the other hand, shows an almost 20\% decrease between 0--$\unit[70]{mM}$ of magnesium. This signifies a decreased tendency of the DNA double helix to unwind under high tensile stress --- in other words, a stabilization of DNA twist. Qualitatively similar results have been shown before in a bulk study of the relaxation of supercoiled, circular dsDNA by topoisomerase I \cite{Xu1997}. There, it was speculated that neutralization of the DNA's backbone charge by magnesium(2+) ions diminishes intramolecular repulsion, effectively stabilizing the helical twist of the molecule. The subsequent inversion of the effect at still higher magnesium concentrations is not observed in our study, however.


In conclusion, we have investigated the effect of magnesium cations on DNA twist--stretch coupling. To this end, we have developed a robust analysis approach for force--extension data, based on the twistable worm-like chain model. Our approach gives access to the elasticity regime between 30 and $\unit[60]{pN}$, and thus to information about twist--stretch coupling, directly from stretching data. We have shown that in the range of magnesium concentrations investigated (0--$\unit[150]{mM}$, in a background of $\unit[500]{mM}$ \chem{NaCl}), the persistence length and stretch modulus of double-stranded DNA are unaffected. DNA twist, however, is stabilized, indicated by a nearly 20\% decrease of the twist--stretch coupling parameter $g_1$.

Interestingly, our analysis also shows that the twistable worm-like chain does not model DNA's full elastic response up until the overstretching transition. A gap of a few piconewton is left, over which the model significantly deviates from the data. This could be due to an early onset of overstretching, possibly in GC-rich areas \cite{King2013}. Future theoretical modeling efforts, possibly taking into account more aspects of DNA's molecular structure, may shed light on this small gap.


\begin{acknowledgments}
We would like to thank T.T. Perkins for suggesting the initial experiment, and M.C.M. de Gunst for useful discussions.
This work has been supported by grants from the Foundation for Fundamental Research on Matter (FOM), which is part of the Netherlands Organization for Scientific Research (NWO), and the European Research Council (consolidator grant).
\end{acknowledgments}

\bibliography{fdfit}

\end{document}